\documentclass[12pt]{article}

\usepackage{amsmath,amsthm,amscd,amssymb,eucal,mathrsfs}
\usepackage{amsfonts}
\usepackage{dsfont}
\usepackage{latexsym}
\usepackage{graphicx}
\usepackage{multicol}

\usepackage[all]{xy}

\begin{document}
\title{ Nonlocal Dynamics of $p$-Adic Strings}

\author{Branko Dragovich \\
Institute of Physics \\ Pregrevica 118, 11080 Zemun,  Belgrade,
Serbia}

\date {~}
\maketitle

\begin{abstract}
We consider the construction of Lagrangians that might  be
suitable for describing  the entire $p$-adic sector of an adelic
open scalar string. These Lagrangians are constructed using the
Lagrangian for $p$-adic strings  with an arbitrary prime number
$p$. They contain space-time nonlocality because of the
d'Alembertian in argument  of the Riemann zeta function. We
present a brief review and some new results.
\end{abstract}

\section{Introduction}

Nonlocal field theory with an infinite number of derivatives has
recently attracted much attention. It is mainly based  on ordinary
and also on $p$-adic string theory, which emerged in 1987
\cite{volovich1}.  Various kinds of $p$-adic strings have been
considered, but the most interesting are strings whose worldsheet
is $p$-adic while all other properties are described by real and
complex numbers. Four-point scattering amplitudes of open scalar
ordinary and $p$-adic strings are connected at the tree level by
their product, which is a constant.  Ordinary and
 $p$-adic strings are treated on an equal footing in this product
(see, e.g. \cite{freund1,volovich2} for a review). Some other
$p$-adic structures have also been  investigated and $p$-adic
mathematical physics was established (see \cite{dragovich} for a
recent review).

Unlike for ordinary strings, there is an effective nonlocal field
theory for the open scalar $p$-adic strings  with a Lagrangian
\cite{freund2,frampton1}  describing four-point scattering
amplitudes and all higher ones at the tree-level. It is worth
noting that this Lagrangian does not contain $p$-adic numbers
explicitly, but only the prime number $p$ which can be regarded as
either a real or a $p$-adic parameter. Because this Lagrangian is
simple and exact at the tree level, it has been essentially used
in the last decade and many aspects of $p$-adic string dynamics
have been considered, compared with the dynamics of ordinary
strings, and applied to nonlocal cosmology (see, e.g.
\cite{sen,zwiebach,vladimirov,arefeva,barnaby} and the references
therein).

This paper contains a review and some new results related to
constructing  a Lagrangian with the Riemann zeta function for the
entire $p$-adic sector of an open scalar string. Requiring of the
Riemann zeta function in the Lagrangian is motivated by the fact
that it appears in  the product over $p$ of all four-point
$p$-adic scalar string amplitudes. In constructing  possible
Lagrangians,  we start from the Lagrangian for a single $p$-adic
open scalar string. An interesting approach to a field theory and
cosmology based on the Riemann zeta function was proposed in
\cite{volovich3}.

\section{Setup: $p$-Adic Open Scalar Strings}

The $p$-adic string theory started analogously to ordinary string
theory with scattering amplitudes. Let $v \in V = \{\infty, \, 2,
\, 3, \, ..., \, p, \, ... \}. $ The crossing symmetric Veneziano
amplitude for scattering of two open scalar strings is defined by
the Gel'fand-Graev-Tate beta function
\begin{equation}
A_v (a, b)  = g_v^2\, \int_{\mathbb{Q}_v} |x|_{v}^{a-1} \,
|1-x|_{v}^{b-1}\, d_v x \,,  \label{1.1}
\end{equation}
where $\mathbb{Q}_p$ is the $p$-adic number field and $a = -
\alpha (s) = - \frac{s}{2} - 1,\, b = - \alpha (t)$ and $ c = -
\alpha (u)$ are complex-valued kinematic variables with the
condition $a + b + c = 1$. We note that the variable $x$ in the
integrands is related to the string worldsheet: the worldsheets of
ordinary and $p$-adic strings are respectively treated by real and
$p$-adic numbers (see, e.g. \cite{freund1,volovich2} and
\cite{gelfand} for the basic properties of $p$-adic numbers and
their functions). Hence $p$-adic strings differ from  ordinary
strings only by the $p$-adic treatment only of the worldsheet.
Integrating in (\ref{1.1}), one obtains
\begin{equation}
A_\infty (a, b)  = g_\infty^2 \frac{\zeta (1-a)}{\zeta (a)}\,
\frac{\zeta (1-b)}{\zeta (b)}\, \frac{\zeta (1-c)}{\zeta (c)}\,,
\label{1.2}
\end{equation}
\begin{equation}
A_p (a, b) = g_p^2\, \frac{1- p^{a-1}}{1- p^{-a}}\, \frac{1-
p^{b-1}}{1- p^{-b}}\,\frac{1- p^{c-1}}{1- p^{-c}}\,, \label{1.3}
\end{equation}
where $\zeta$ is the Riemann zeta function. Expression (\ref{1.2})
is for the ordinary case and (\ref{1.3}) is for the $p$-adic case.

The Riemann zeta function
\begin{equation}
\zeta (s) = \sum_{n= 1}^{+\infty} \frac{1}{n^{s}} = \prod_p
\frac{1}{ 1 - p^{- s}}\,, \quad s = \sigma + i \tau \,, \quad
\sigma
>1\,, \label{1.4}
\end{equation}
 has an analytic continuation to the entire complex-$s$ plane
excluding the point $s=1$, where it has a simple pole with unit
residue.   Taking the product of $p$-adic string amplitudes
(\ref{1.3}) over $p$ and using (\ref{1.4}), we obtain (see, e.g.
\cite{dragovich8})
%\begin{equation}
%\prod_p A_p (a, b) =  \frac{\zeta (a)}{\zeta (1-a)}\, \frac{\zeta
%(b)}{\zeta (1-b)}\, \frac{\zeta (c)}{\zeta (1-c)} \, \prod_p
%g_p^2\,, \label{1.5}
%\end{equation}
%that gives a  simple  formula
\begin{equation}
 \prod_v A_v (a, b) =  \prod_v g_v^2\, = \text{const}. \label{1.6}
\end{equation}
The product of  $p$-adic amplitudes in (\ref{1.6})   diverges
\cite{dragovich1}, but it  converges after an appropriate
regularization. Requiring that amplitude product  (\ref{1.6}) be
finite implies that the product of coupling constants is finite,
i.e. $g_\infty^2 \, \prod_p g_p^2 = const.$ There are three
interesting possibilities: (i)\, $g_p^2 =1$, \, (ii)\, $g_p^2 =
\frac{p^2}{p^2 - 1}$, which gives $\prod_p g_p^2 = \zeta (2)$, \,
and (iii)\, $g_p^2 = |\frac{m}{n}|_p$, where $m$ and $n$ are any
two nonzero integers, and this gives $g_\infty^2 \, \prod_p g_p^2
= |\frac{m}{n}|_\infty \prod_p |\frac{m}{n}|_p = 1 $.

It follows from (\ref{1.6})  that the ordinary Veneziano
amplitude, which is a special function, can be expressed as the
product of all inverse $p$-adic counterparts, which are elementary
functions. This is  a consequence of the Gel'fand-Graev-Tate beta
functions and  is not a general property of string scattering
amplitudes. In the general case, the string amplitude product  is
a function of kinematic variables.

Another interpretation of expression (\ref{1.6}) is related to an
adelic string. But  an adelic string should have an adelic
worldsheet.  A scattering amplitude of two open scalar strings
with their adelic worldsheets has not yet been obtained.
Therefore, the concept of an adelic string with an adelic
worldsheet is not  well founded and remains questionable. But
$p$-adic strings with a $p$-adic worldsheet are well defined, and
the string amplitude product  for open scalar strings has a useful
meaning.

The exact tree-level Lagrangian of the effective scalar field
$\varphi$, which describes the open $p$-adic string tachyon, is
\cite{freund2,frampton1}
\begin{equation} {\cal L}_p = \frac{m^D}{g_p^2}\, \frac{p^2}{p-1} \Big[
-\frac{1}{2}\, \varphi \, p^{-\frac{\Box}{2 m^2}} \, \varphi  +
\frac{1}{p+1}\, \varphi^{p+1} \Big]\,,  \label{2.1} \end{equation}
where $p$
 is a prime, $\Box = - \partial_t^2  + \nabla^2$ is the
$D$-dimensional d'Alembertian. The corresponding equation of
motion for (\ref{2.1})
%\begin{equation} p^{-\frac{\Box}{2 m^2}}\, \varphi = \varphi^p \,,
%\label{2.4} \end{equation} %and it
has been investigated by many authors (see, e.g. \cite{vladimirov}
and the references therein).

We now want to consider construction of Lagrangians that can be
used to describe entire $p$-adic sector of an  open scalar string.
In particular, an appropriate such Lagrangian should describe the
scattering amplitude, which contains the Riemann zeta function.
Consequently, this Lagrangian must contain the Riemann zeta
function with the d'Alembertian in its argument. We should
therefore seek possible constructions of a Lagrangian that
contains the Riemann zeta function and is closely related to
$p$-adic Lagrangian (\ref{2.1}). There are additive and
multiplicative approaches; we mainly consider  the additive
approach below.

\section{Additive and Multiplicative Approaches}

The prime number $p$ in (\ref{2.1})  can be replaced by any
natural number $n \geq 2$ and the results make sense. We introduce
a Lagrangian  incorporating all  Lagrangians (\ref{2.1}) described
above but with $p$ replaced by $n \in \mathbb{N}$. The
corresponding sum of all Lagrangians ${\cal L}_n$  is
\begin{equation} L =   \sum_{n = 1}^{+\infty} C_n\, {\cal L}_n   =
m^D \sum_{n= 1}^{+\infty}  \frac{ C_n}{g_n^2}\frac{n^2}{n -1}
\Big[ -\frac{1}{2}\, \phi \, n^{-\frac{\Box}{2 m^2}} \, \phi +
\frac{1}{n + 1} \, \phi^{n+1} \Big]\,, \label{3.1}
\end{equation} whose concrete form depends on the
choice of the coefficients $C_n$ and coupling constants $g_n$. We
set $$\frac{C_n}{g_n^2}\frac{n^2}{n-1} = D_n , \, \, n = 1, 2,
...$$. The following simple cases lead to the Riemann zeta
function: $D_n = 1, \, D_n = (-1)^{n-1}, \,  D_n = n+1, \, D_n =
\mu(n),  \, D_n = - \mu(n) (n+1),$ and  $ D_n = (-1)^{n-1} (n+1)$,
where $\mu (n)$ is the M\"obius function.

The case $D_n = 1$ was considered in \cite{dragovich3,dragovich4}
%Obtained Lagrangian is
%\begin{equation} L =  m^D \Big[ \,- \frac{1}{2}\,
% \phi \,  \zeta\Big(\frac{\Box}{2 \, m^2}\Big) \, \phi    + {\cal{AC}} \sum_{n= 1}^{+\infty} \frac{\phi^{n+1}}{n + 1} \,
%  \Big]\,, \label{3.2} \end{equation} where
%$\mathcal{AC}$ denotes analytic continuation.
and the case $ D_n = n+1$ was investigated in \cite{dragovich5}.

%and the corresponding Lagrangian is

 %\begin{equation} L =  m^D \Big[ \, - \frac{1}{2}\,
 %\phi \,  \Big\{ \zeta\Big({\frac{\Box}{2\, m^2}  -
 %1}\Big)\, + \, \zeta\Big({\frac{\Box}{2\, m^2} }\Big) \Big\} \, \phi \,  + \,   \frac{\phi^2}{1 - \phi} \,
 %\Big]\,. \label{3.3} \end{equation}

The variants with the M\"obius function $\mu (n)$ are described in
\cite{dragovich6} and \cite{dragovich7}.  We recall that its
explicit definition is
\begin{equation}
\mu (n)= \left \{ \begin{array}{lll} 0 , \quad &  n = p^2 m \,, \\
(-1)^k , \quad & n = p_1 p_2 \cdots p_k ,\,\,  p_i \neq p_j \,, \\
1 , \quad & n = 1, \,\,  (k=0)\, ,
\end{array} \right.
\label{3.4}
\end{equation}
and it is related to the inverse Riemann zeta function by
\begin{equation}
\frac{1}{\zeta (s)} = \sum_{n =1}^{+\infty}\, \frac{\mu (n)}{n^s},
\quad s=\sigma + i \tau , \quad \sigma > 1\,. \label{3.5}
\end{equation}

The corresponding Lagrangian for $D_n = \mu (n)$ is
\begin{equation}
L =  m^D \Big[ - \frac{1}{2}\, \phi \, \frac{1}{
\zeta\Big({\frac{\Box}{2 m^2}}\Big)} \,\phi + \int_0^\phi {\cal
M}(\phi) \, d\phi\Big] , \label{3.6}
\end{equation}
where ${\cal M}(\phi) = \sum_{n= 1}^{+\infty} {\mu (n)} \,
\phi^{n} = \phi - \phi^2 - \phi^3 - \phi^5 + \phi^6 - \phi^7 +
\phi^{10} - \phi^{11} - \dots $.

For $D_n = - \mu (n)\, (n+1)$  the Lagrangian is
\begin{equation}
{ L} = m^D\,  \Big\{ \frac{1}{2} \, \phi \Big[ \frac{1}{\zeta
\Big( \frac{\Box}{2 m^2} - 1 \Big)} + \frac{1}{\zeta \Big(
\frac{\Box}{2 m^2}  \Big)}\Big] \, \phi - \phi^2 \, F(\phi)\Big\}
\,, \label{3.7}
\end{equation}
where $F (\phi)\, = \, \sum_{n=1}^{+\infty}\, \mu (n) \phi^{n-1} =
\, 1 - \phi - \phi^2 - \phi^4 +...$.

The case  with $D_n = (-1)^{n-1}\, (n+1)$  was recently introduced
 in \cite{dragovich8}. We recall that
\begin{equation}
\sum_{n= 1}^{+\infty} (-1)^{n-1} \frac{1}{n^{s}} =  (1 - 2^{1-s})
\, \zeta (s), \quad s = \sigma + i \tau \,, \quad \sigma
> 0\,, \label{3.8}
\end{equation}
which has  an analytic continuation to the entire complex-$s$
plane without singularities, i.e. the analytic expression
\cite{wikipedia} is
\begin{equation}
(1 - 2^{1-s}) \, \zeta (s) = \sum_{n=0}^\infty \frac{1}{2^{n+1}}
\, \sum_{k=0}^n (-1)^k \, \left(\begin{array}{c}
n\\
k
\end{array}
\right) (k +1)^{-s} \,. \label{3.8a}
\end{equation}
 At point $s = 1$, one has $\lim_{s\to 1} (1 - 2^{1-s}) \, \zeta
(s)\, = \, \sum_{n= 1}^{+\infty} (-1)^{n-1} \frac{1}{n} \, = \,
\log 2$. Applying (\ref{3.8}) to (\ref{3.1}) and using analytic
continuation we obtain
 \begin{align} \nonumber L = & - m^D \Big[ \,  \frac{1}{2}\,
 \phi \,  \Big\{ \, \Big(1 - 2^{2 - \frac{\Box}{2 m^2}}\Big)\, \zeta\Big({\frac{\Box}{2\, m^2}  -
 1}\Big)\, \\ & + \,  \Big(1 - 2^{1 - \frac{\Box}{2 m^2}}\Big)\, \zeta\Big({\frac{\Box}{2\, m^2} }\Big)
 \Big\} \, \phi \,  - \,   \frac{\phi^2}{1 + \phi} \,
 \Big]\,. \label{3.9} \end{align}

We now consider the case $D_n = (-1)^{n-1}$. The corresponding
Lagrangian is
\begin{align}
L = m^D \, \Big[ - \frac{1}{2} \phi \Big( 1 - 2^{1- \frac{\Box}{2
m^2}} \Big) \zeta\Big(\frac{\Box}{2 m^2}\Big) \phi \, + \phi -
\frac{1}{2} \,\log(1 + \phi)^2 \Big] . \label{3.14}
\end{align}
The potential is
\begin{align}
V(\phi) = - L (\Box = 0) = m^D \Big[ \frac{1}{4} \, \phi^2 - \phi
+ \frac{1}{2} \log (1 +\phi)^2 \Big] ,   \label{3.15}
\end{align}
which has one local maximum $V(0) = 0$ and one local minimum at
$\phi = 1$. It is singular at $\phi = -1$, i.e. $V(-1) = -
\infty$, and $V (\pm \infty) = + \infty$.  The equation of motion
is
\begin{align}
\Big( 1 - 2^{1- \frac{\Box}{2 m^2}} \Big) \zeta\Big(\frac{\Box}{2
m^2}\Big) \phi = \frac{\phi}{1+ \phi} , \label{3.16}
\end{align}
which has  the two trivial solutions: $\phi = 0$ and $\phi = 1$.

%\subsection{Multiplicative approach}

 The Riemann zeta function arising in the multiplicative  approach is given
in the form of product (\ref{1.4}). The initial Lagrangian is
$p$-adic Lagrangian (\ref{2.1}) with $g_p^2 = \frac{p^2}{p^2 -1}$.
%We rewrite (\ref{2.1}) in the form
%\begin{align} \nonumber {\cal L}_p = & m^D\,  \Big\{
%\frac{1}{2}\, \varphi \, \Big[ \Big(1 - p^{-\frac{\Box}{2 m^2}+1}
%\Big) + \Big( 1 - p^{-\frac{\Box}{2 m^2}}\Big) \Big]\, \varphi \\
%& - \varphi^2 \Big( 1 - \varphi^{p-1} \Big) \Big\}\,. \label{3.11}
%\end{align}
%Taking products
%\begin{equation}
% \prod_p (1 - p^{-\frac{\Box}{2 m^2}+1}) \,,
% \quad \prod_p (1 - p^{-\frac{\Box}{2 m^2}}) \,, \quad \prod_p ( 1 - \varphi^{p-1})  \label{3.12}
%\end{equation} in (\ref{3.11}) at the relevant places one obtains
%Lagrangian
%\begin{align}
%{\mathcal L} =  \frac{m^D}{g^2}\,  \Big\{ \frac{1}{2} \, \phi
%\Big[\frac{1}{\zeta \Big( \frac{\Box}{2 m^2} - 1 \Big)} +
%\frac{1}{\zeta \Big( \frac{\Box}{2 m^2}  \Big)}\Big] \, \phi -
%\phi^2 \, \Phi (\phi) \Big\} \,, \label{3.13}
%\end{align}
%where $\Phi (\phi) = \prod_p \Big( 1 - \phi^{p-1} \Big) = 1 - \phi
%-\phi^2 + \phi^3 - \phi^4 + ... $.
The Lagrangian  obtained in this approach \cite{dragovich7}  is
similar to  (\ref{3.7}) above. These two Lagrangians describe the
same field theory in the week field approximation.

%Lagrangian (\ref{3.13}) was considered in \cite{dragovich7}.

\section{Concluding remarks}

In the preceding section, we  presented  some Lagrangians that can
be used  to describe the $p$-adic sector of open scalar strings.
They contain the Riemann zeta function and  are also starting
points for interesting examples of the so-call zeta field theory.
The corresponding potentials, which are $V (\phi) = - L (\Box =
0)$, and equations of motions are considered in the cited
references. All these zeta field theory models contain tachyons.

The most interesting  of the above Lagrangians are  (\ref{3.9})
and (\ref{3.14}). Unlike the other Lagrangians, these have no
singularity with respect to the d'Alembertian $\Box$, and it is
easier to apply the pseudodifferential treatment.  This
analyticity of the Lagrangian is expected to be useful in its
application to nonlocal cosmology, in particular, using
linearization procedure (see, e.g., \cite{koshelev} and references
therein).

%We would like also to point out Lagrangians (\ref{3.7}) and
%(\ref{3.13}), since they are mutually very similar. These
%Lagrangians describe the same model at the weak field
%approximation, although they are constructed using rather
%different approaches.

\section*{\large Acknowledgements}
The paper was supported in part by the Ministry of Science and
Technological Development, Serbia (Contract No. 144032D). The
author thanks organizers of the International Bogolyubov
Conference ``Problems of Theoretical and Mathematical Physics''
(August 21-27, 2009, Moscow-Dubna, Russia) for a very pleasant and
useful scientific meeting.

 \end{document}